\documentclass[10pt,english,oneside,twocolumn,a4paper]{article}

\makeatletter
\let\blx@rerun@biber\relax
\makeatother

\usepackage{graphicx}
\usepackage[utf8]{inputenc}
\usepackage{mathptmx}
\usepackage{tabularx}
\usepackage{ragged2e}
\usepackage[singlelinecheck=false]{caption}
\usepackage[T1]{fontenc}
\usepackage{amsmath}
\RequirePackage[blocks]{authblk}
\usepackage{babel}
\usepackage{xcolor}%
\definecolor{todonotecol}{RGB}{250,0,0}%
\usepackage{verbatim}
\usepackage{textcomp} 
\usepackage{pgfplots}
\usepackage{tikz}
\usepackage[%
	acronym,%
	nopostdot,%
	seeautonumberlist,%
	shortcuts,%
	section=chapter,%
	toc,%
]{glossaries}%
\glsaddkey*{url}
{}
{\glsentryurl}
{\Glsentryurl}
{\glsurl}
{\Glsurl}
{\GLSurl}
\glsaddkey*{longpltwo}
{}
{\glsentrylongpltwo}
{\Glsentrylongpltwo}
{\glslongpltwo}
{\Glslongpltwo}
{\GLSlongpltwo}
\newglossaryentry{}{%
	name={},%
	plural={},%
	description={},%
}%
\newacronym{dmz}{DMZ}{demilitarized zone}
\newacronym{it}{IT}{Information Technology}
\newacronym{ot}{OT}{operational technology}
\newacronym{opcua}{OPC UA}{Open Platform Communications Unified Architecture}
\newacronym{tsn}{TSN}{Time-Sensitive Networking}
\newacronym{scada}{SCADA}{supervisory control and data acquisition}
\newacronym{vpc}{VPC}{Virtual Process Controller}
\newacronym{plc}{PLC}{programmable logic controller}
\newacronym{udp}{UDP}{User Datagram Protocol}
\newacronym{pubsub}{Pub/Sub}{Publish/Subscribe}
\newacronym{vm}{VM}{Virtual Machine}
\newacronym{pc}{PC}{personal computer}
\newacronym{ptp}{PTP}{Precision Time Protocol}
\newacronym{ntp}{NTP}{Network Time Protocol}
\newacronym{splc}{Soft-PLC}{Software-PLC}
\newacronym{rte}{RTE}{Real-Time Ethernet}
\newacronym{osi}{OSI}{Open Systems Interconnection}
\newacronym{mac}{MAC}{Media Access Control}
\newacronym{e2e}{E2E}{end-to-end}
\newacronym{tcp}{TCP}{Transmission Control Protocol}
\newacronym{hmi}{HMI}{Human-Machine Interface}
\newacronym{kpi}{KPI}{Key Performance Indicator}
\newacronym{mes}{MES}{Manufacturing Execution System}
\newacronym{rami4.0}{RAMI 4.0}{Reference Architectural Model Industrie 4.0}
\newacronym{tacnet4.0}{TACNET 4.0}{Tactile Internet 4.0}
\newacronym{3gpp}{3GPP}{3rd Generation Partnership Project}
\newacronym{5gppp}{5GPPP}{5G Infrastructure Public Private Partnership}
\newacronym{itu}{ITU}{International Telecommunication Union}
\newacronym{ngnm}{NGNM}{Next Generation Mobile Networks}
\newacronym{agv}{AGV}{automated guided vehicle}
\newacronym{v2v}{V2V}{vehicle-to-vehicle}
\newacronym{v2x}{V2X}{vehicle-to-infrastructure}
\newacronym{d2x}{D2X}{device-to-x}
\newacronym{qos}{QoS}{Quality of Service}
\newacronym{noa}{NOA}{NAMUR Open Architecture}
\newacronym{dcs}{DCS}{distributed control system}
\newacronym{m2m}{M2M}{machine-to-machine}
\newacronym{sdn}{SDN}{software-defined networking}
\newacronym{erp}{ERP}{enterprise resource planning}
\newacronym{ip}{IP}{Internet Protocol}
\newacronym{lte}{LTE}{Long Term Evolution}
\newacronym{5g}{5G}{5th generation wireless communication system}
\newacronym{6g}{6G}{6th Generation Wireless Communication Systems}
\newacronym{harq}{HARQ}{Hybrid Automatic Repeat Request}
\newacronym{rat}{RAT}{radio access technology}
\newacronym{mc}{MC}{Multi-Connectivity}
\newacronym{fbmc}{FBMC}{Filter Bank Multicarrier}
\newacronym{gfdm}{GFDM}{Generalized Frequency Division Multiplexing}
\newacronym{sla}{SLA}{service-level agreement}
\newacronym{5qi}{5QI}{5G QoS Indicator}
\newacronym{bmbf}{BMBF}{German Federal Ministry of Education and Research}
\newacronym{nrt}{NRT}{non real-time}
\newacronym{irt}{IRT}{isochronous real-time}
\newacronym{rt}{RT}{real-time}
\newacronym{ar}{AR}{augmented reality}
\newacronym{wlan}{WLAN}{wireless local area network}
\newacronym{cpu}{CPU}{central processing unit}
\newacronym{ram}{RAM}{random-access memory}
\newacronym{ue}{UE}{User Equipment}
\newacronym{ran}{RAN}{radio access network}
\newacronym{bs}{BS}{base station}
\newacronym{cn}{CN}{Core Network}
\newacronym{epc}{EPC}{evolved packet core}
\newacronym{mec}{MEC}{mobile edge cloud}
\newacronym{rtt}{RTT}{Round-Trip Time}
\newacronym{vpn}{VPN}{virtual private network}
\newacronym{vpf}{VPF}{Virtual Process Control Function}
\newacronym{vdi}{VDI}{Association of German Engineers}
\newacronym{ipc}{IPC}{industrial pc}
\newacronym{rtos}{RTOS}{real-time operating system}
\newacronym{vpcmo}{VPC M\&O}{VPC Management \& Orchestration}
\newacronym{i/o}{I/O}{input /output}
\newacronym{fbd}{FBD}{Function Block Diagram}
\newacronym{ldd}{LDD}{Ladder Logic Diagram}
\newacronym{sfc}{SFC}{Sequential Function Chart}
\newacronym{il}{IL}{Instruction List}
\newacronym{st}{ST}{Structured Text}
\newacronym{os}{OS}{Operating System}
\newacronym{ir}{IR}{Inactive Resource}
\newacronym{msc}{MSC}{message sequence chart}
\newacronym{icps}{ICPS}{industrial cyber-physical system}
\newacronym{cps}{CPS}{cyber-physical systems}
\newacronym{mtd}{MTD}{Moving Target Defense}
\newacronym{iot}{IoT}{Internet of Things}
\newacronym{iiot}{IIoT}{industrial Internet of Things}
\newacronym{iiot2}{IIoT}{Industrial IoT}
\newacronym{k8s}{K8s}{Kubernetes}
\newacronym{rkt}{rkt}{CoreOS Rocket}
\newacronym{dcos}{DC/OS}{Distributed Cloud Operating System}
\newacronym{ml}{ML}{Machine Learning}
\newacronym{ai}{AI}{Artificial Intelligence}
\newacronym{ict}{ICT}{industrial communication technology}
\newacronym{kvm}{KVM}{kernel-based virtual machine}
\newacronym{ie}{IE}{Industrial Ethernet}
\newacronym{ias}{IAS}{industrial automation system}
\newacronym{cots}{COTS}{commercial off-the-shelf}
\newacronym{slat}{SLAT}{Second Level Address Translation}
\newacronym{gui}{GUI}{graphical user interface}
\newacronym{l2}{L2}{Layer 2}
\newacronym{l3}{L3}{Layer 3}
\newacronym{sme}{SME}{small and medium-sized enterprise}
\newacronym{sba}{SBA}{Service Based Architecture}
\newacronym{xaas}{XaaS}{Everything-as-a-Service}
\newacronym{5gc}{5GC}{5G Core}
\newacronym{vnf}{VNF}{Virtual Network Function}
\newacronym{nfv}{NFV}{Network Functions Virtualization}
\newacronym{amf}{AMF}{Access and Mobility Management Function}
\newacronym{ng-ran}{NG-RAN}{Next Generation Radio Access Network}
\newacronym{lmf}{LMF}{Location management function}
\newacronym{pcf}{PCF}{Policy Control Function}
\newacronym{ausf}{AUSF}{Authentication Server Function}
\newacronym{udm}{UDM}{Unified Data Management}
\newacronym{upf}{UPF}{User Plane Function}
\newacronym{smf}{SMF}{Session Management Function}
\newacronym{udr}{UDR}{Unified Data Repository}
\newacronym{nef}{NEF}{Network Exposure Function}
\newacronym{nf}{NF}{Network Function}
\newacronym{nrf}{NRF}{Network Repository Function}
\newacronym{mano}{MANO}{Management and Orchestration}
\newacronym{c/r}{C/R}{Checkpoint/Restore}
\newacronym{criu}{CRIU}{Checkpoint/Restore In Userspace}
\newacronym{ppm}{PPM}{Parallel Process Migration}
\newacronym{pdu}{PDU}{Protocol Data Unit}

\glsdisablehyper

\usepackage[
    backend=biber,
    style=ieee,
    language=english,
   bibwarn=true,
]{biblatex}

\DefineBibliographyStrings{english}{
   andothers = {{et\,al\adddot}},
}
\makeatletter
\let\ps@plain\ps@empty
\def\@xivpt{14bp}

\def\@sect#1#2#3#4#5#6[#7]#8{%
  \ifnum #2>\c@secnumdepth
    \let\@svsec\@empty
  \else
    \refstepcounter{#1}%
    \protected@edef\@svsec{%
      \ifnum #2<4
        \hb@xt@10mm{\csname the#1\endcsname}\relax
      \else
        \hb@xt@12mm{\csname the#1\endcsname}\relax
      \fi}%
  \fi
  \@tempskipa #5\relax
  \ifdim \@tempskipa>\z@
    \begingroup
      #6{%
        \@hangfrom{\hskip #3\relax\@svsec}%
          \interlinepenalty \@M #8\@@par}%
    \endgroup
    \csname #1mark\endcsname{#7}%
    \addcontentsline{toc}{#1}{%
      \ifnum #2>\c@secnumdepth \else
        \protect\numberline{\csname the#1\endcsname}%
      \fi
      #7}%
  \else
    \def\@svsechd{%
      #6{\hskip #3\relax
      \@svsec #8}%
      \csname #1mark\endcsname{#7}%
      \addcontentsline{toc}{#1}{%
        \ifnum #2>\c@secnumdepth \else
          \protect\numberline{\csname the#1\endcsname}%
        \fi
        #7}}%
  \fi
  \@xsect{#5}}
\renewcommand\LARGE{\@setfontsize\LARGE{16}{20}}
\def\abstract#1{\def\@abstract{#1}}
\def\abstractEn#1{\def\@abstractEn{#1}}
\def\titleEn#1{\def\@titleEn{#1}}
\def\@maketitle{%
  \newpage
  \null
  \let \footnote \thanks
    {\LARGE\bfseries\RaggedRight \@titleEn \par}%
    \vskip 1\baselineskip%
    {\normalsize
      \@author\par}%
    \vskip 2\baselineskip%
    {\section*{Abstract}
      \@abstractEn}%
  \par
  \vskip 3\baselineskip
  }
\renewcommand\section{\@startsection {section}{1}{\z@}%
                                   {-3.5ex \@plus -1ex \@minus -.2ex}%
                                   {\baselineskip}%
                                   {\normalfont\Large\bfseries\RaggedRight}}
\renewcommand\subsection{\@startsection{subsection}{2}{\z@}%
                                     {\baselineskip}%
                                     {1ex}%
                                     {\normalfont\large\bfseries\RaggedRight}}
\renewcommand\subsubsection{\@startsection{subsubsection}{3}{\z@}%
                                     {1\baselineskip}%
                                     {3bp}%
                                     {\normalfont\normalsize\bfseries\RaggedRight}}
\renewcommand\paragraph{\@startsection{paragraph}{4}{\z@}%
                                    {1\baselineskip\@plus1ex \@minus.2ex}%
                                    {3bp}%
                                    {\normalfont\normalsize\RaggedRight}}
\renewcommand\subparagraph{\@startsection{subparagraph}{5}{\parindent}%
                                       {3.25ex \@plus1ex \@minus .2ex}%
                                       {-1em}%
                                      {\normalfont\normalsize\bfseries\RaggedRight}}
\parindent\p@
\makeatother
\DeclareCaptionLabelSeparator{enskip}{\enskip}
\title{Towards Organic 6G Networks: Virtualization and Live Migration of Core Network Functions}
\titleEn{Towards Organic 6G Networks: Virtualization and Live Migration of Core Network Functions\thanks{This research was supported by the German Federal Ministry of Education and Research (BMBF) within the Open6GHub project under grant number 16KISK003K. The responsibility for this publication lies with the authors. This is a preprint of a work accepted but not yet published at the Mobile Communication-Technologies and Applications; 25. ITG-Symposium. Please cite as: M. Gundall, J. Stegmann, C. Huber, and H.D. Schotten: “Towards Organic 6G Networks: Virtualization and Live Migration of Core Network Functions”. In: Mobile Communication-Technologies and Applications; 25. ITG-Symposium, VDE, 2021.}}
\author[1]{Michael Gundall}
\author[1]{Julius Stegmann}
\author[1]{Christopher Huber}
\author[12]{Hans D. Schotten}
\affil[1]{Intelligent Networks, German Research Center for Artificial Intelligence GmbH (DFKI), Trippstadter Straße 122, 67663 Kaiserslautern, Germany, Email: \{michael.gundall, julius\_raphael.stegmann, christopher.huber, hans\_dieter.schotten\}@dfki.de}
\affil[2]{Institute for Wireless Communication and Navigation, University of Kaiserslautern (TUK), Gottlieb-Daimler Straße 47, 67663 Kaiserslautern, Germany, Email: schotten@eit.uni-kl.de}

\abstractEn{
In the context of Industry 4.0, more and more mobile use cases are appearing on industrial factory floors. These use cases place high demands on various quantitative requirements, such as latency, availability, and more. In addition, qualitative requirements such as flexibility are arising. Since virtualization technology is a key enabler for the flexibility that is required by novel use cases and on the way to organic networking as it is addressed by \gls{6g}, we investigate container virtualization technology in this paper. We focus on container technology since \gls{os}-level virtualization has multiple benefits compared to hardware virtualization, such as \glspl{vm}. \newline
Thus, we discuss several aspects of container based virtualization, e.g. selection of suitable network drivers and orchestration tools, with respect to most important \gls{5gc} functions. In addition, the functions have different quantitative or qualitative requirements depending on whether they are stateless or stateful, and whether the specific function is located at either the control or user plane. Therefore, we also analyze the aforementioned live migration concepts for the \gls{5gc} functions and evaluate them based on well-defined metrics, such as migration time and process downtime.}
				\newcommand{\disablewr}[1]{#1}%
				\newcommand{\newcommanddisw}[3]{\newcommand{#1}[1]{\disablewr{\textcolor{#2}{#3}}}}%
\definecolor{todocol}{named}{red}
\newcommanddisw{\todo}{todocol}{ToDo: #1}%
\definecolor{migucol}{named}{purple}%
\newcommanddisw{\migucom}{migucol}{{@}comment: #1}%
\newcommanddisw{\miguhigh}{migucol}{#1}%
\begin{document}
\maketitle

\section{Introduction}

In the scope of Industry 4.0, more and more mobile use cases appear in industrial factory halls \cite{etfa2018, etfa2021}. These use cases have stringent demands on different requirements, such as latency, availability, and more. Therefore, high performance wireless communications systems are required. Here, mobile radio communications, such as 5G \cite{access2021,Mobilkom2019} and 6G \cite{jiang2021road}, can play an important role. Besides the aforementioned quantitative requirements, there are also qualitative requirements that raise novel challenges and opportunities. Examples for these requirements are security, integration possibilities, and flexibility. 

\begin{figure}[h!]
\centering
  \includegraphics[scale=.9]{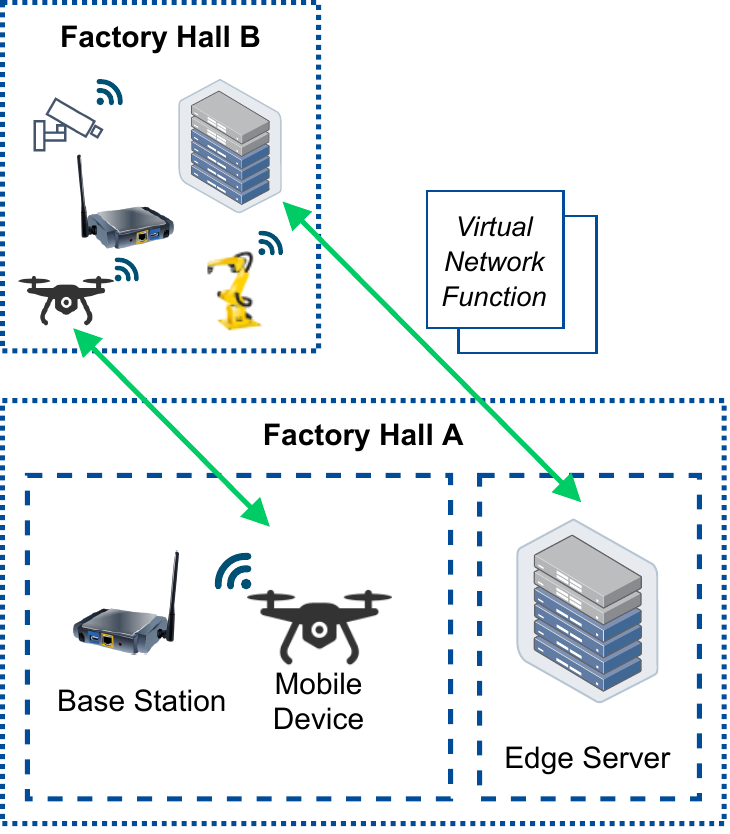}
  \caption{Exemplary industrial use case, where a drone moves between factory halls.}
  \label{fig: use case}
\end{figure}

Therefore, Fig. \ref{fig: use case} shows and exemplary use case that requires both low-latency communication as well as a high flexibility. 
If a mobile device, such as a drone, offloads certain algorithms, it is important that this algorithm is executed by an edge server that is located as close as possible to this device. If the drone moves between factory halls or even factories the algorithm has to be processed by another server. Besides the required flexibility on application side, also communication networks have to support this mobility. In order to deliver data packets in time, several network functions have to be deployed close to the mobile device. Here, the so-called \gls{nfv} comes into place. Together with virtualization technologies, such as \gls{os}-level virtualization and hardware virtualization it is possible to automatically deploy and run \glspl{vnf} on nearly any device that offers computational resources. Thus, we investigate, whether existing technologies are suitable for the application of \gls{nfv} for functions of the \gls{5gc} in industrial environments. 

Therefore, the paper is structured as follows: Sec.~\ref{sec:Related Work} gives an overview about related work on this topic, while Sec.~\ref{sec:Key Technologies for the Realization of Organic Networking} presents key technologies for the realization of organic networking. Moreover, Sec.~\ref{sec:5G SBA} details 5G \gls{sba} in detail and introduces both chances and challenges given by virtualization and live migration for relevant \gls{5gc} functions. Finally, a conclusion is given (Sec.~\ref{sec:Conclusion}).


\section{Related Work}
\label{sec:Related Work} 

In order to achieve the flexibility that is demanded by emerging mobile use cases, virtualization technology can be used, whereas hardware and \gls{os}-level virtualization are well-known concepts in the \gls{it} environment. Thus, it has been shown that \gls{os}-level virtualization using Linux containers is more efficient compared to traditional \glspl{vm} that belong to hardware virtualization  \cite{7164727, 10.1145/2851613.2851737, indin2020}. Furthermore, the authors in \cite{10.1145/2851613.2851737,indin2020} investigated the use of \gls{os}-level virtualization technology for industrial applications. Even if both works are targeted for industrial automation systems, the results can be transferred to \glspl{vnf} of \gls{5gc}, since they place comparable requirements.


In order to improve flexibility, 5G applies the \gls{sba} paradigm. Consequently, the functions are not only service-based but also more fine grained, compared to ealier technologies, such as 4G. Due to this reason, it can be assumed that the application of virtualization technologies to \gls{5gc} is advantageous compared 4G, even if there are also approaches for applying \gls{xaas} to 4G \glspl{cn} \cite{taleb2015ease}.

\section{Key Technologies for the Realization of Organic Networking}
\label{sec:Key Technologies for the Realization of Organic Networking} 

In order to realize organic networking, several technologies, which are well-known in the \gls{it}, have to be introduced in the communication domain. Therefore, this section introduces related technologies and concepts.

\subsection{Container Virtualization}
As already mentioned, several works indicate that virtualization using containers is suitable if efficiency and performance of the \gls{vnf} are important \cite{7164727, 10.1145/2851613.2851737, indin2020}. Here, the network drivers play a central role. However, they differ not only in performance, but also in their networking capabilities and security level, such as network isolation.

Thus, Tab. \ref{tab:netwrok drivers} gives an overview about the standard network drivers of Docker containers regarding \gls{rtt}, which was measured between containers that were deployed on two different hosts, networking capabilities, and security level.
\begin{table}[h!]
\caption{Network driver overview \cite{indin2020}.}
\begin{center}
\begin{tabularx}{\columnwidth}{|c|c|c|c|}
\hline 
 \textbf{Netw. Driver} & \textbf{RTT [µs]} & \textbf{Networking} & \textbf{~Security~}  \\
\hline 
Host  & 522 & L2 / L3 &  - \\
Bridge & 600 & L2 / L3 & $\circ$ \\
Macvlan & 520 & L2 / L3  &  $\circ$ \\
Ipvlan (L2)  & 520 &  L3 & $\circ$ \\
Ipvlan (L3)   & 539 &  L3& $\circ$  \\
  \hline
 Overlay  & 656  &  (L2)\textsuperscript{1}/L3 &  + \\
  \hline
  \multicolumn{4}{l}{\textsuperscript{1} Only valid for L2 overlay network drivers of K8s.}
\end{tabularx}
\label{tab:netwrok drivers}
\end{center}
\end{table}

While efficiency and performance, such \gls{rtt} and overhead, could be most important for several applications \cite{reichardt2021benchmarking}, some industrial applications require special networking capabilities, such as \gls{l2} support, which means the exchange of Ethernet frames without \gls{ip} layer (\gls{l3}). A typical example for this are \gls{ie} protocols and \gls{tsn}. Since this feature is not supported by all Docker network drivers by a rule, it is also a selection criteria that should be considered.

\subsection{Container Orchestration}
If an automated deployment and scaling of a service is required, an orchestration tool, such as Docker Swarm or \gls{k8s}, is required. Here, it is important to name that they typically bring up additional network drivers that build overlay networks. In case of Docker Swarm, the "Overlay" network driver is not able to transmit \gls{l2} packets, while \gls{k8s} has several \gls{l2} overlay network drivers, e.g. multus. However, for Docker Swarm it is possible to use several standard network drivers of Docker also for a scalable service but requires more configuration effort.

Furthermore, both orchestration tools allow to automatically deploy services and to create as much replicas as required. This method can be used for load balancing as well as for the application of fail-over mechanisms.  Here, \gls{k8s} provides more possibilities to create highly individualized and complex service compositions that are called "Deployment". The reason for this is probably the higher industry support \cite{bernstein2014containers}.

\subsection{Live Migration Approaches}
The aforementioned service composition can typically only be applied in order to replicate containers that are not state synchronized. On the other hand, if a stateful container should be redeployed, e.g., due to mobility requirements, live migration is a possible method. Thus, the so-called \gls{c/r} tactic has become widely accepted for the live migration of processes. Here, a process is "frozen" and its current status on the disk is checkpointed. This data can then be transferred to a new target system. There, the process can be restarted at exactly the same point in time at which it was previously frozen. In the last few years, developments have increasingly been moving in the direction of user-space-based methods. These offer the enormous advantage of high transparency in combination with not too invasive intervention in the central components of the operating system. 

The Linux Foundation introduced its \gls{criu} software in 2012 and has since further developed it into a powerful tool for live migration of processes. In the meantime, \gls{criu} is either integrated into OpenVZ, LXC/LXD, Docker, and Podman or can be used in combination with them without much effort \cite{criu1}. While live migration with \gls{criu} is already widespread in the area of high-performance computing \cite{hpc1}, its use in other application areas has been rather limited so far. 

\begin{figure}[h!]
\centering
  \includegraphics[width=\columnwidth]{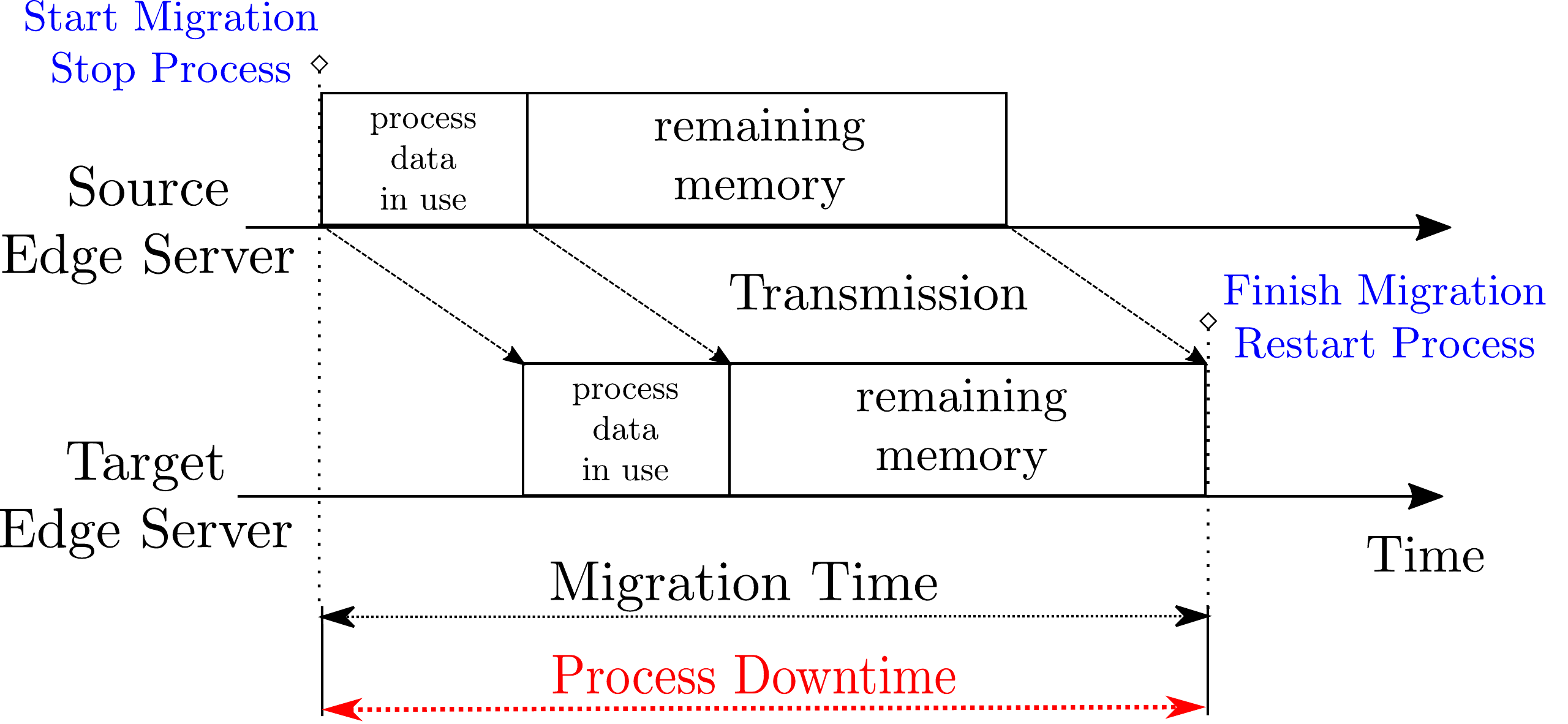} 
  \caption{\gls{c/r} migration with inter-copy memory transfer.}
  \label{fig: inter copy}
\end{figure}

The main focus of research here is on memory transfer, which is indispensable for process migration. In a classical (inter-copy) \gls{c/r} procedure, which is shown in Fig. \ref{fig: inter copy}, 
the process is frozen, all data in the memory is completely transferred from one system to another, before the process is restarted. The downtime of the process and the migration time are therefore almost identical. To further minimize the downtime, two primary strategies can be used: pre- and post-copying of the memory. 

\begin{figure}[h!]
\centering
\includegraphics[width=\columnwidth]{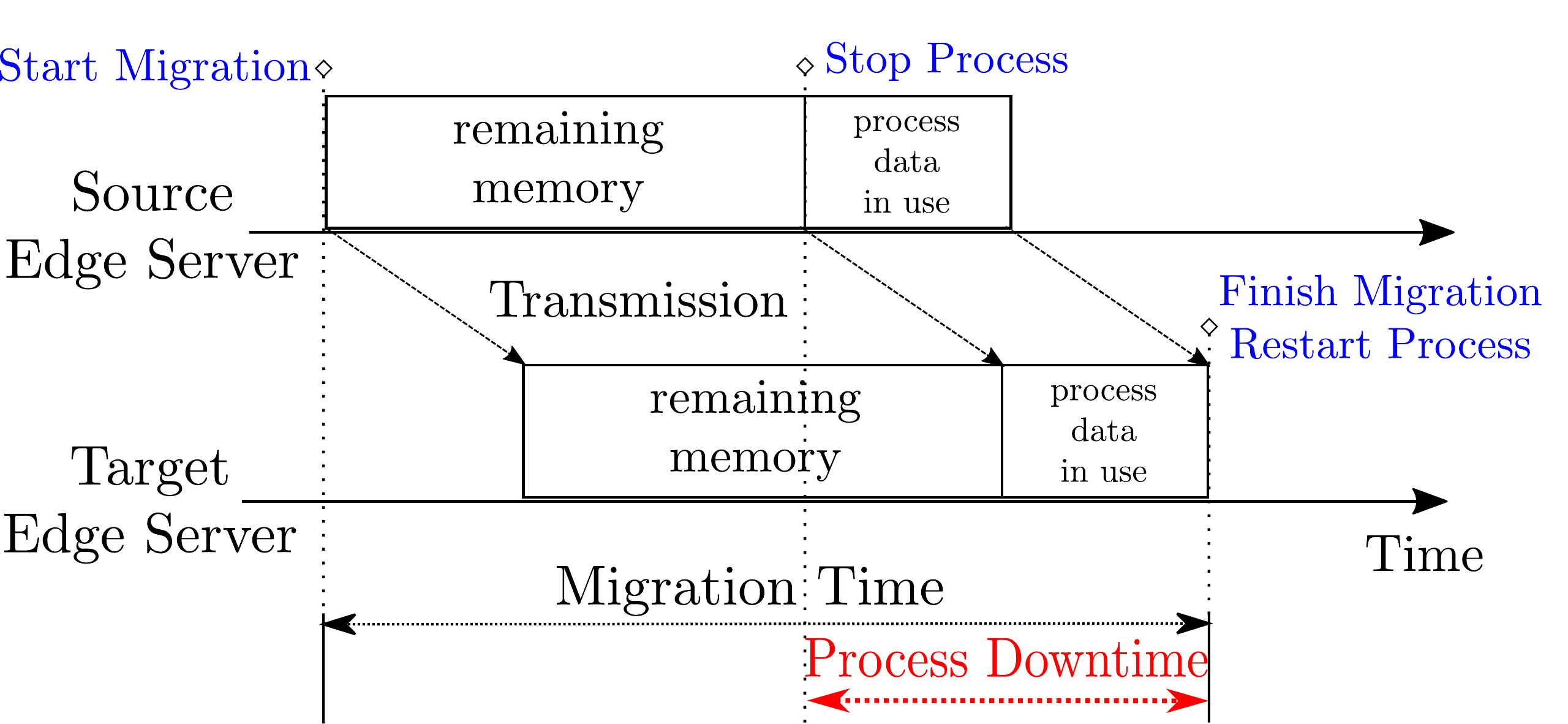} 
  \caption{\gls{c/r} migration with pre-copy memory transfer.}
  \label{fig: pre copy}
\end{figure}

\begin{figure}[h!]
\centering
  \includegraphics[width=\columnwidth]{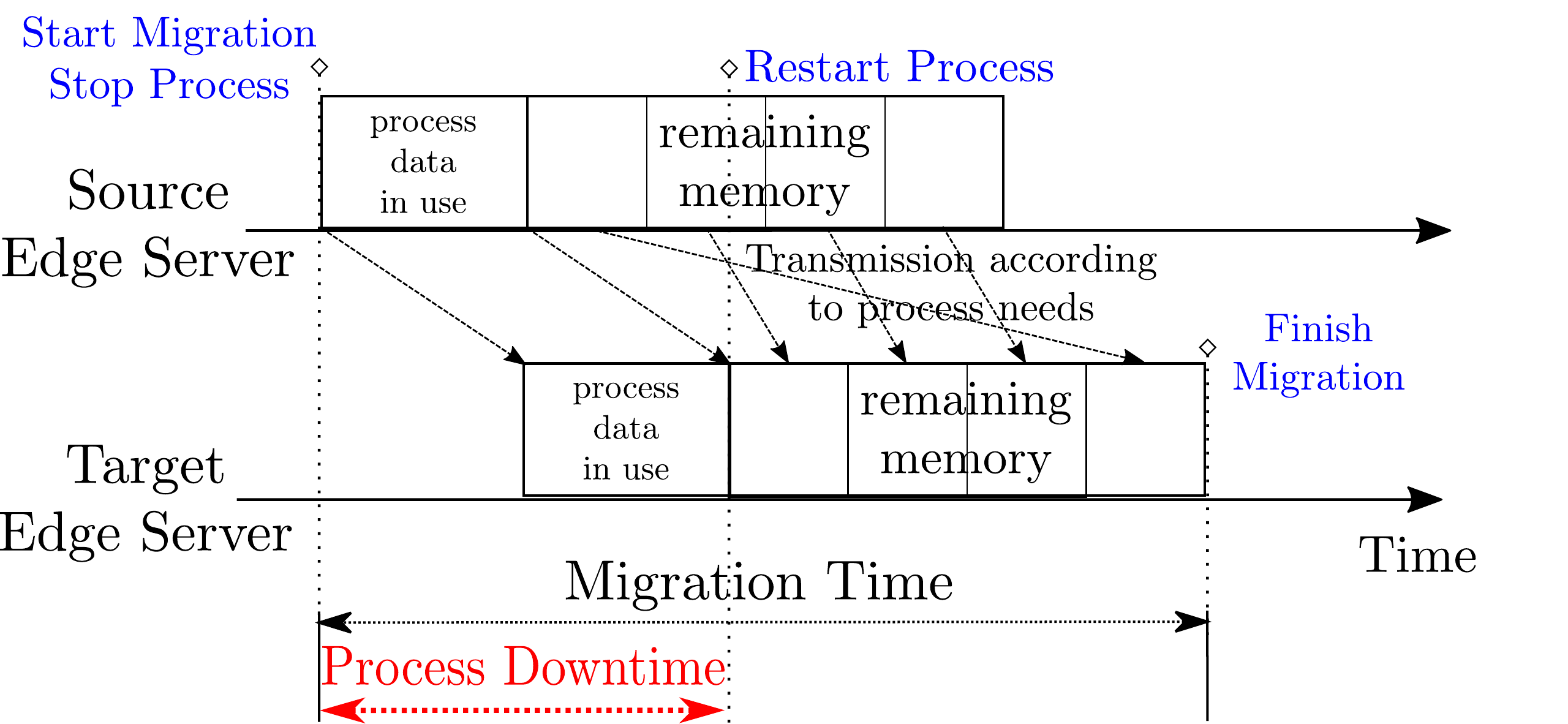} 
  \caption{\gls{c/r} migration with post-copy memory transfer.}
  \label{fig: poat copy}
\end{figure}

In the pre-copy tactic (see Fig. \ref{fig: pre copy}), 
as much data as possible is first transferred to the target system, primarily data that is not expected to be needed in the coming process iterations. Then the process is shut down on the source system, the remaining data is transferred, and the process is restarted on the target system. With the post-copy tactic (see Fig. \ref{fig: poat copy}), 
on the other hand, the process is frozen immediately at the start of the migration process, similar to the inter-copy method. Afterwards, however, only the parts of the memory that are important for the next process iterations are transferred. The remaining parts of the memory are then transferred after the process has already restarted on the target system \cite{reber1}. 

Both strategies are part of intensive research \cite{performance1,precopy1}. The post-copy strategies in particular increase the risk of a complete process failure if missing data cannot be transferred in time afterwards. The pre-copy strategy brings few advantages in terms of downtime if large parts of the data change in just a few process steps. Both methods require additional precise prediction of future steps. 

\begin{figure}[h!]
\centering
  \includegraphics[width=\columnwidth]{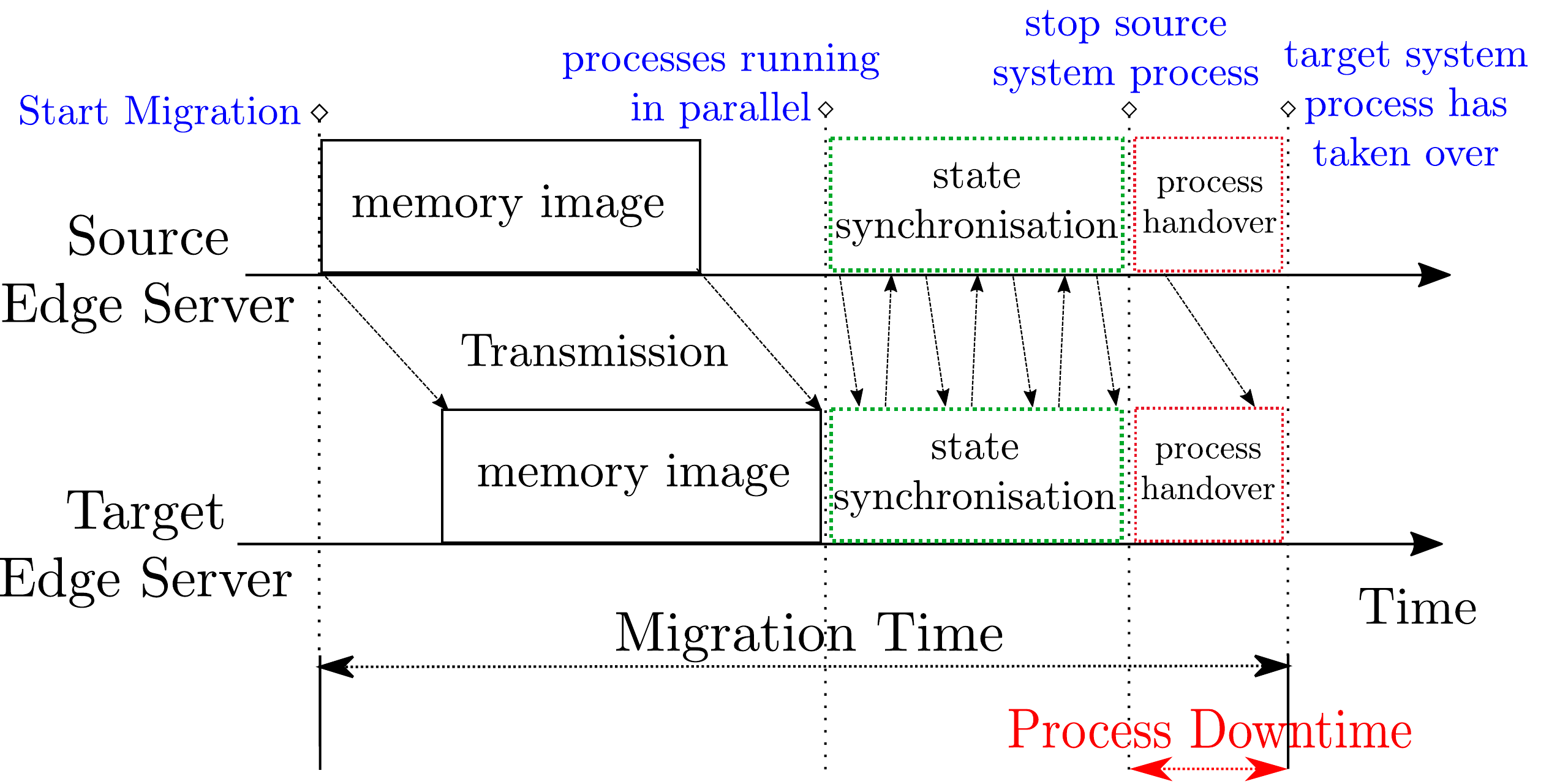} 
  \caption{\gls{ppm} procedure including handover mechanism.}
  \label{fig: parallel process}
\end{figure}

Therefore, latest approaches go one step further and use the \gls{ppm} methodology \cite{parallel1,parallel2}. In previous approaches, only one instance of the process was active at a time. Thus, Fig. \ref{fig: parallel process} 
depicts the idea that the process is already running on the target system and both processes are supplied with the same data. If a migration is triggered, ideally only a very small part of the memory still has to be transferred to the target system. This leads in a considerably reduced downtime. However, there are multiple challenges that lie on the one hand in managing a smooth handover, such as time and state synchronization, and on the other hand in checking that all instances of the processes running in parallel are always supplied with the identical data at the same time.

\section{5G \acrfull{sba}}
\label{sec:5G SBA} 

\begin{figure*}[h!]
\centering
  \includegraphics[scale=1.0, clip]{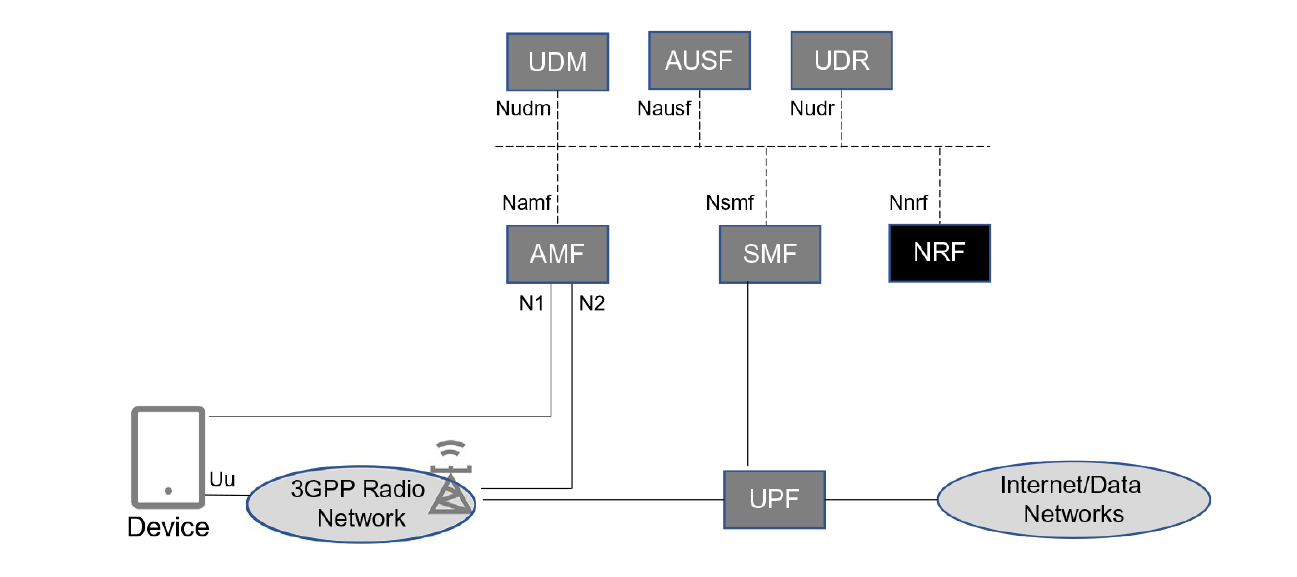}
  \caption{Mandatory components of a 5G network architecture and corresponding interfaces \cite{rommer20195g}.}  \label{fig1}
\end{figure*}

This section introduces the 5G \gls{sba} and discusses the possibilities and challenges of organic networking for most relevant \gls{5gc} functions. Therefore, Fig. \ref{fig1} shows the mandatory components of a 5G system and their corresponding interfaces. 

Furthermore, the functions that are explained in the following sections (\mbox{Sec. \ref{subsec: upf}-\ref{subsec: nrf}}), can be mapped to either the user plane or control plane. While user plane traffic is most important for end user applications, control plane contains the relevant functions for a suitable operation of the 5G system. Therefore, a decrease in \gls{qos} in the user plane has a direct impact in end user applications, while performance variations in the control plane do not necessarily affect the end user application. 

\subsection{\gls{upf}}
\label{subsec: upf} 

The main task of the \gls{upf}, which is located in the user plane, is the processing and forwarding of user data, with the \gls{smf} controlling its functionality. This implies that the \gls{upf} can be considered stateless, but has high demands on latency and availability, since a failure would cause a direct loss of connectivity for end users. It connects to external \gls{ip} networks serving as an anchor for the \gls{ue} towards the external network, hiding the mobility. As a result, \gls{ip} packets with a destination address belonging to a \gls{ue} are always routed from the Internet to the specific \gls{upf} serving that device, regardless of whether the device is moving around the network. The \gls{upf} generates records of charging data and traffic usage which can be sent to the \gls{smf}. It also performs packet inspections that can be used for applying configured policies, gating, redirecting traffic, and applying data rate limits. In addition, it can also apply \gls{qos} polices in the downlink direction. 
Additionally, 5G systems allow not only the possibility for \gls{ip} based \gls{pdu} Sessions, but support also Ethernet \gls{pdu} Session type \cite{rommer20195g,3gpp.23.501}. 

Since the \gls{upf} is stateless, live migration is not required. However, it is suitable to use virtualization technology in order to automatically deploy and restart \glspl{upf} on each targeted hardware node. Moreover, multiple instances of \glspl{upf} can be deployed on one device, e.g., to apply redundancy or load balancing mechanisms. Since \gls{k8s} has benefits regarding deployment policies, this orchestration tool can be the preferred option for this function. However, if all \gls{pdu} Session types should be supported, the standard network driver of \gls{k8s} cannot be used and a specialized third party network driver is required, in order to transmit \gls{l2} data packets. Alternatively Docker Swarm in combination with one of the standard Docker network drivers could be an appropriate solution.

\subsection{\acrfull{smf}}
\label{subsec: smf} 
The \gls{smf}, which is part of control plane, is mainly responsible for the management of the end user sessions. The main tasks are creating, updating and deleting \gls{pdu} Sessions, and managing the session context with \gls{upf}. It communicates indirectly with end user devices through the \gls{amf}, which forwards session-related messages between the devices and the \glspl{smf}. Separating other control plane functions from the user plane, the \gls{smf} takes over some of the functions previously performed by the MME and assumes the role of DHCP server and \gls{ip} address management system. Additionally, the \gls{smf} plays a crucial role in the charging-related function within the network. By collecting its own charging data, it manages the charging functions of the \gls{upf}. 

As already indicated, the \gls{smf} is stateful. Thus, live migration approaches should be applied if this function should be redeployed on a different hardware node. This can be required, e.g., if the hardware node is more close to the \gls{ue}, and very fast and dynamic reconfigurations of the corresponding \glspl{upf} are required, as it is the case for mobile devices that have high demands on latency and are covering a wide serving area. If a high service availability should be guaranteed, pre-copy \gls{c/r} migration or \gls{ppm} are suitable live migration approaches.

\subsection{\acrfull{amf}}
\label{subsec: amf} 
The \gls{amf} is responsible for the interaction between the \gls{ng-ran} via the N2 interface as well as with the interaction between \gls{ue} via the N1 interface. The \gls{amf} is part of most signaling call flows in a 5G network, providing support for encrypted signaling connections to devices in order to register, authenticate, and switch between different radio cells in the network. It is also responsible for paging \glspl{ue} in the idle state. The \gls{amf} relays all session management related signals between \gls{amf} and \gls{ue}, which is different from the 4G \gls{cn} architecture. A further difference consists in the fact that \gls{amf} itself does not perform authentication, but orders it as a service from \gls{ausf} \cite{3gpp.23.501}.

Due to the fact that all control layer data flows between \gls{ue} and \gls{5gc} as well as \gls{ng-ran} and \gls{5gc} are forwarded by the \gls{amf} to other \glspl{nf}, e.g., \gls{smf}, the requirements on service availability are even higher compared to \gls{smf}. Therefore, the application of \gls{ppm} can be the preferred live migration approach.

\subsection{\acrfull{ausf}}
\label{subsec: ausf} 
The \gls{ausf} functions are rather limited, but very important. It provides the authentication service of a specific \gls{ue} using the authentication data created by \gls{udm}, as well providing services that allow secure updating of roaming information and other parameters in the \gls{ue}.

Since the \gls{ausf} is highly security relevant, it should not be compromised by an attacker. Therefore, both network and guest/host isolation should be high for this function. Here, overlay networks can be superior compared to other network drivers. Since a service outage would only prevent novel devices to join the network, no special needs for latency and service availability are required. Thus, inter-copy  migration is the best option for live migration, since it minimizes the migration time and overhead of the process, because all data has only to be send once. However, the cases where a live migration of the \gls{ausf} is required seems quite limited.

\subsection{\acrfull{udm}}
\label{subsec: udm} 
The \gls{udm} manages data for access authorization,  data network profiles, and user registration, all of which are managed by the \gls{smf}. In addition, access is authorized for specific users based on subscription data. For instance, for roaming subscribers and home subscribers, this could mean that different access rules apply. \gls{udm} can be stateful or stateless \cite{3gpp.29.503}. In case of a stateful version, data is stored locally, whereas a stateless version stores the data externally in the \gls{udr}. With a stateful architecture, data is shared between services that manage the communication between network layers. The disadvantage is that in case of a problem, all services that are sharing information must be taken down from the network at once. With a stateless architecture, subscriber data is kept separate from the functions that support it. This provides more stability and flexibility because database access is separate from the operational network, but also prevents the same information from being updated at the same time by multiple nodes, which can cause delays in the network. With more than one instance of \gls{amf} and \gls{smf} in the network, the \gls{udm} keeps track of which instance is serving a particular device.

In case of the stateful version it is most important that the states are transferred correctly. Since a small service downtime should not cause direct loss of connectivity, traditional inter-copy \gls{c/r} migration is sufficient. Additionally, no synchronization error or similar could occur. In the stateless version, either \gls{k8s} or Docker Swarm orchestration tool can be used, since no special needs on networking performance or capabilities are given. However, in this case, the \gls{udr} is stateful, and inter-copy \gls{c/r} migration can be applied for this function.

\subsection{\acrfull{udr}}
\label{subsec: udr} 
The \gls{udr} is the central storage where the structured data is stored. For instance, the \gls{udm} can store and retrieve subscriber data such as access and mobility data or network slice selection data. Equally, the \gls{pcf} can store policy-related data or the \gls{nef} can store structured data for exposure and application data. Multiple \gls{udr} systems may be deployed in the network, each taking different data sets or subsets, or serving different \glspl{nf}.

\subsection{\acrfull{nrf}}
\label{subsec: nrf} 
The \gls{nrf} is one of the most important components of the 5G architecture. It provides a single record of all \glspl{nf}, along with the services provided by each element that can be instantiated, scaled and terminated without or minimal manual intervention in the operator's network.

The \gls{nrf} places equal demands on virtualization and live migration as \gls{udm}/\gls{udr}. However, the migration time and the corresponding downtime might be higher, dependent on its size and the data amount that has to be transferred. In this case, it has to be carried out if either process downtime or migration time should be minimized. If the migration time is most important, C/R migration with inter-copy memory transfer can be used. Otherwise, pre-copy \gls{c/r} or \gls{ppm} is beneficial.

\section{Conclusion}
\label{sec:Conclusion}
In this paper, we investigated key technologies that are required by organic networking that is targeted by \gls{6g}. Therefore, we proposed the recent state of research for both virtualization and live migration technologies. Additionally, we introduced most important \gls{5gc} functions and analyzed them based on latency and availability requirements.

\printbibliography
\end{document}